\documentclass[11pt,a4paper]{article}
\usepackage{amsmath,amssymb}
\usepackage{epsfig,graphicx}

\def\beq{\begin{equation}}
\def\eeq{\end{equation}}
\def\beqn{\begin{eqnarray}}
\def\eeqn{\end{eqnarray}}

\begin{document}

\title{\bf BFKL Pomeron and Bern-Dixon-Smirnov amplitudes in
$N=4$ SUSY \footnote{Talk at the International Conference "Quarks-08",
May 2008, Sergiev Pasad, Russia}}
\author{L.N.Lipatov\\
\small{\em Petersburg Nuclear Physics Institute, Russia} \\
\small{\em Hamburg University, Germany}
}
\date{}
\maketitle

\begin{abstract}
We
review the theoretical
approaches for investigations
of the high energy
hadron-hadron scattering in the
Regge kinematics.
It is demonstrated, that
the gluon in QCD is
reggeized and the
Pomeron is a composite
state of the reggeized gluons.
Remarkable properties
of the BFKL equation for
the Pomeron wave function
in QCD and supersymmetric
gauge theories are outlined.
Due to the
AdS/CFT correspondence
the BFKL Pomeron is equivalent
to the reggeized graviton in
the extended N=4 SUSY.
The properties of the
maximal transcendentality
and integrability are realized
in this model. The BDS multi-gluon
scattering amplitudes are investigated
in the Regge limit. They 
do not contain
the Mandelstam cuts and are not valid beyond one loop. 
It is shown, that
the hamiltonian for these composite states
coincides with the hamiltonian of
an integrable open Heisenberg spin chain.
\end{abstract}

\section{High energy interactions}
Hadron-hadron scattering
in the Regge kinematics 
\beq
s=(p_A+p_B)^2 =(2E)^2 >> \vec{q}^2=-(p_{A'}-p_A)^2 \sim m^2
\eeq
is usually described in terms of a $t$-channel exchange of the
Reggeon
(see Fig.1)

\beq
A_p(s,t)=\xi _p(t)\,g(t)\,s^{j_p(t)}\,g(t)\,,\,\,j_p(t)=j_0+\alpha   
't\,,\,\,\xi_p(t)=\frac{e^{-i\pi j_p(t)}+p}{\sin (\pi j_p)}\,,
\eeq
where $j_p(t)$ is the Regge trajectory which is assumed to be linear,
$j_0$ and $\alpha '$ are its
itercept and slope, respectively. The signature factor $\xi_p$ is a
complex quantity depending on the Reggeon signature $p=\pm 1$. 
\begin{figure}[ht]
\vspace{0cm} \par
\begin{center}
\leavevmode
\epsfysize=2cm
 \epsfxsize=2cm \epsffile{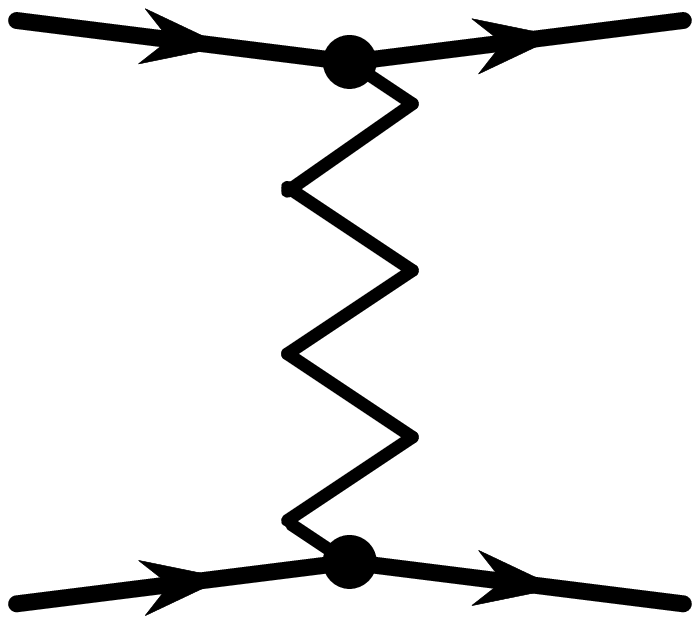}
\end{center}
\par
\vspace{0cm} 
\label{Reggeon}
\end{figure}

A special Reggeon
- Pomeron with vacuum quantum numbers is introduced to explain 
an approximately constant behavior
of total cross-sections at high energies and a fullfillment of the Pomeranchuck
theorem $\sigma _{h\bar{h}}/\sigma _{hh}\rightarrow 1$. Its signature
$p$ is positive and its intercept is
close to unity
$j_0^p=1+\Delta \,,\,\,\Delta \ll 1$.

The particle production at high energies can be investigated
in the multi-Regge kinematics (see Fig.2)
\begin{figure}[ht]
\vspace{0cm} \par
\begin{center}
\leavevmode
\epsfysize=4cm
 \epsfxsize=4cm \epsffile{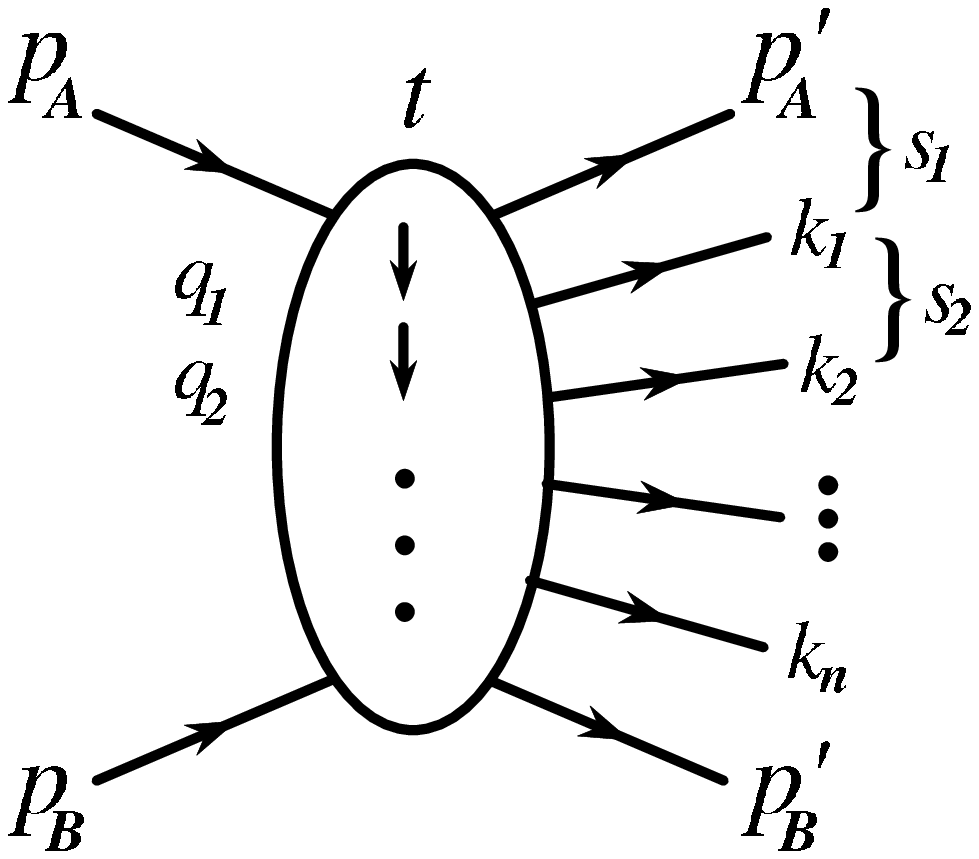}
\end{center}  
\par
\vspace{0cm} 
\label{MultRegg}
\end{figure}
\beq
s\gg s_1\,,\,\,s_2\,,...\,,\,\,s_{n+1}\gg
t_1\,,\,\,t_2\,,...\,,\,\,t_{n+1}\,,
\eeq
where $s_r$ are squares of the sum of neibouring particle momenta
$k_{r-1},\,k_r$ and $-t_r$ are squares of the momentum transfers
$\vec{q}_r$. The production amplitude in the framework of the Regge
model is also expressed in terms of the
Reggeon exchanges in
each of $t_r$-channels
\beq
A_{2\rightarrow 2+n}\sim \prod_{r=1}^{n+1}s_r^{j_p(t_r)}\,,
\eeq
where we neglected the signature factors, which will be discussed later.

\section{ \bf Gluon reggeization in QCD}

In the Born approximation of QCD the scattering amplitude
for two colored particle scattering is factorized 
\beq
M_{AB}^{A^{\prime }B^{\prime }}(s,t)|_{Born}=\Gamma _{A^{\prime
}A}^{c}\,\frac{2s}{t}\,\Gamma _{B^{\prime }B}^{c}\,,\,\,
\Gamma _{A^{\prime }A}^{c}
=g\,T_{A^{\prime }A}^{c}\,\delta _{\lambda
_{A^{\prime }}\lambda _{A}}\,,
\eeq
where $T^c$ are the generators of the color group $SU(N_c)$
in the corresponding
representation and $\lambda _r$ are helicities of the
colliding and final state particles. In the leading logarithmic
approximation (LLA)
the scattering amplitude in QCD can be written as follows~\cite{BFKL}
\beq
M_{AB}^{A^{\prime }B^{\prime }}(s,t)=M_{AB}^{A^{\prime
}B^{\prime }}(s,t)|_{Born}\,s^{\omega (t)},\,\,\alpha _{s}\ln s
\sim 1 \,,
\eeq
where the gluon Regge trajectory is

\beq
\omega (-|q|^2)=-\frac{\alpha _s N_c}{(2\pi ) ^{2-2\epsilon}}\,|q^2|\int
\frac{\mu^{2\epsilon}\,d^{2-2\epsilon}k}{|k|^2|q-k|^2}
 \approx
-a\,\left(\ln \frac{|q^2|}{\mu ^2}-\frac{1}{\epsilon}\right)\,.
\eeq
Here the extra dimensions $2\epsilon \rightarrow -0$ were introduced
to regularize the infrared divergency, the parameter $\mu$ is 
the renormalization point and we used the notation
\[
a=\frac{\alpha _s\,N_c}{2\pi}\,\left(4\pi\,e^{-\gamma}\right)^\epsilon\,,
\]
where $\gamma$ is the Euler constant $\gamma =-\psi (1)$.
This Regge trajectory
was calculated also in two-loop approximation in QCD~\cite{trajQCD} and in
supersymmetric gauge theories~\cite{trajN4}.

Further, the gluon production amplitude in the multi-Regge
kinematics at LLA can be written in the 
factorized form~\cite{BFKL} (see Fig.2)  
\beq
M_{2\rightarrow 1+n}
=2s\,\Gamma _{A^{\prime
}A}^{c_1}\, \frac{s_1^{\omega _1}}{|q_1|^2}
\,gT_{c_2c_1}^{d_1}C(q_2,q_1)\, \frac{s_2^{\omega
_2}}{|q_2|^2}...C(q_{n},q_{n-1})\, \frac{s_{n}^{\omega
_{n}}}{|q_{n}|^2}\,\Gamma _{B^{\prime
}B}^{c_n}\,.
\eeq
The Reggeon-Reggeon-gluon vertex for the produced gluon
with a definite helicity is
\beq
C(q_2,q_1)=\frac{q_2\,q_1^*}{q_2^*-q_1^*}\,,
\eeq
where we used the complex notations for the transverse components
of particle momenta.

It gives a possibility to calculate the total
cross-section~\cite{BFKL}
\beq
\sigma _t=\sum _n
\int d\Gamma _n\left|M_{2\rightarrow 1+n}\right|^2\,,
\eeq
where $\Gamma _n$ is the phase space for the produced particle
momenta in the multi-Regge kinematics.

\section{BFKL equation}
Because the production amplitudes
in QCD are   
factorized, one can write a Bethe-Salpeter-type equation
for the total cross-section $\sigma _t$. Using
also the optical
theorem it
can be presented as the equation of
Balitsky, Fadin, Kuraev and Lipatov (BFKL) for the Pomeron
wave function~\cite{BFKL}

\beq
E\,\Psi (\vec{\rho}_{1},\vec{\rho}_{2})=H_{12}\,\Psi (\vec{\rho}_{1},\vec{%
\rho}_{2})\;,\,\,\Delta =-\frac{\alpha
_{s}N_{c}}{2\pi }\,E\,,
\eeq
where $\sigma _t \sim s^\Delta _{max}$ and the BFKL Hamiltonian
in the coordinate representation $\rho$ is
\beq
H_{12}=\ln \,|p_1p_2|^{2}+
\frac{1}{p_1p_2^*}(\ln |\rho _{12}|^{2})p_1p_2^*
+\frac{1}{p_1^*p_2}(\ln |\rho _{12}|^{2})p_1^*p_2
-4\psi (1)
\eeq
and $\rho _{12}=\rho _1-\rho _2$.
It is invariant under the M\"obius transformations~\cite{int1, moeb}
\beq
{\large \rho _k \rightarrow \frac{a\rho _k+b}{c\rho _k+d}}
\eeq
and has the property of the holomorphic separability
\beq
H_{12}=h_{12}+h^*_{12}\,,\,\,h_{12}=\ln (p_1p_2)+
\frac{1}{p_1}\ln (\rho _{12}) \,p_1
+\frac{1}{p_2}\ln (\rho _{12}) \,p_2
-2\psi (1)\,.
\label{pairh}
\eeq
Here we used the complex notations for two-dimensional 
transverse coordinates and
their canonically conjugated momenta.
The conformal weights for the principal series of
unitary representations of the M\"{o}bius group are
\beq
m=\gamma +n/2\,,\,\,\widetilde{m}=\gamma -n/2\,,\,\,\gamma
=1/2+i\nu \,,
\eeq
where $\gamma$ is the anomalous dimension of the twist-2 operators
and $n$ is conformal spin.

The Bartels-Kwiecinski-Praszalowicz (BKP) equation
for colorless  composite states of several reggeized gluons has the following
form~\cite{BKP}

\beq
E\,\Psi (\vec{\rho}_{1},...)=
H\,\Psi
(\vec{\rho}_{1},...)\;,\,\,H=\sum _{k<l}
\frac{\vec{T}_k\vec{T}_l}{-N_c}\,H_{kl}\,,
\eeq
where $H_{kl}$ is the BFKL hamiltonian.
Apart from the M\"{o}bius invariance its wave function
in the multi-color QCD ($N_c \rightarrow \infty$) has the property of the
holomorphic factorization~\cite{separ}

\beq
\Psi (\vec{\rho} _1,...,\vec{\rho}_n)=\sum _{r,s}
a_{r,s}\,\Psi _r(\rho _1,...,\rho _n)\,
\Psi _s (\rho _1^*,...,\rho _n^*)\,,
\eeq
where the sum is performed over a degenerate set of solutions for the corresponding
holomorphic and anti-holomorphic equations. The BKP equation has the
duality symmetry $p_k\rightarrow \rho _{k,k+1}\rightarrow p_{k+1}$
($k=1,2,...,n$)~\cite{dual} and $n$
integrals of motion $q_r,\,q_r^*$~\cite{int}. The corresponding hamiltonians
$h$ and $h^*$ are local hamiltonians of the integrable Heisenberg spin model, in which
spins are generators of the M\"{o}biuos group~\cite{LiFK}. We can introduce
the transfer ($T$) and monodromy ($t$) matrices according to the definitions~\cite{int}
\beq
T(u)=tr\,t(u)=\sum_{r=0}^{n}u^{n-r}\,q_{r}\,,\,\,t(u)=
L_{1}L_{2}...L_{n}\,,
\eeq
\beq
L_{k}=\left(
\begin{array}{cc}
u+\rho _{k}\,p_{k} & p_{k} \\
-\rho _{k}^{2}\,p_{k} & u-\rho _{k}\,p_{k}\end{array}\right)\,,\,\,
t(u)=\left(
\begin{array}{cc}
A(u) & B(u) \\
C(u) & D(u)\end{array}\right)\,.
\label{ABCD}
\eeq
The matrix elements of $t(u)$ satisfy some bilinear commutation relations
following from the
Yang-Baxter equation~\cite{int}
\beq
t_{r_{1}^
{\prime
}}^{s_{1}}(u)\,t_{r_{2}^{\prime
}}^{s_{2}}(v)\,l_{r_{1}r_{2}}^{r_{1}^{\prime }r_{2}^{\prime
}}(v-u)=l_{s_{1}^{\prime }s_{2}^{\prime
}}^{s_{1}s_{2}}(v-u)\,t_{r_{2}}^{s_{2}^{\prime
}}(v)\,t_{r_{1}}^{s_{1}^{\prime }}(u)\,,\,\,\hat{l}(u)=u\,
\hat{1}+i\,\hat{P}\,,
\eeq
where $\hat{l}(u)$ is the monodromy matrix for the usual Heisenberg spin model
and $\hat{P}$ is the permutation operator.
This equation can be solved with the use of the Bethe ansatz and the
Baxter-Sklyanin approach~{\cite{Veg, DKM}}.

\section{Pomeron in $N=4$ SUSY}

One can calculate the integral kernel for the BFKL equation
also in
two loops~\cite{FL}. Its eigenvalue can be written as follows
\beq
\omega =4\,\hat{a}\,\,\chi (n,\gamma )+4\,\,\hat{a}^{2}\,\Delta
(n,\gamma )\,,\,\,\hat{a}=g^{2}N_{c}/(16\pi ^{2})\,,
\eeq
where
\beq
 \chi
(n,\gamma )=2\Psi (1)-\Psi (\gamma +|n|/2)-\Psi (1-\gamma +|n|/2)
\eeq
and $\Psi (x)=\Gamma '(x)/\Gamma (x)$. The one-loop correction
$\Delta (n, \gamma )$ in QCD contains the non-analytic terms -
the Kroniker symbols
$\delta _{|n|,0}$ and $\delta _{|n|,2}$~\cite{trajN4}.
But in $N=4$ SUSY they are cancelled and
we obtain for $\Delta (n, \gamma )$ the following result   
in
the hermitially separable form~\cite{trajN4, KL}
\beq
\Delta (n,\gamma )=\phi (M)+\phi (M^{\ast })-\frac{
\rho (M)+ \rho (M^{\ast })}{2\hat{a}/\omega}\,,\,M=\gamma
+\frac{|n|}{2}\,,
\eeq 
\beq
\rho (M)=\beta ^{\prime
}(M)+\frac{1}{2}%
\zeta (2)\,,\,\beta ^{\prime }(z)=\frac{1}{4}\Biggl[\Psi ^{\prime
}\Bigl(\frac{z+1}{2}\Bigr)-\Psi ^{\prime
}\Bigl(\frac{z}{2}\Bigr)\Biggr]\,.
\eeq
It is interesting, that all functions entering in these expressions
have the property of the maximal transcendentality~\cite{KL}.
In particular,
$\phi (M)$ can be written in the form
\beq
\phi (M)=3\zeta (3)+\Psi ^{^{\prime \prime }}(M)-2\Phi (M)+ 2\beta
^{^{\prime }}(M)\Bigl(\Psi (1)-\Psi (M)\Bigr),
\eeq
\beq
\Phi (M)=\sum_{k=0}^{\infty } \frac{(-1)^{k}}{k+M}\left( \Psi ^{\prime
}(k+1)~-
~\frac{%
\Psi (k+1)-\Psi (1)}{k+M}\right)\,.
\eeq
Here $\Psi (M)$ has the transcedentality equal to 1, its derivatives
$\Psi ^{(n)}$ have transcedentalities $n+1$, the additional poles in
the sum over $k$ increase the transcedentality of the
function $\Phi (M)$ up to 3 being also the transcendentality of 
$\zeta (3)$.
The maximal transcendentality
hypothesis is
valid also for the anomalous dimensions of twist-2 -operators in
$N=4$ SUSY~\cite{KLV, KLOV} contrary to the case of QCD~\cite{VMV}.

The stationary BFKL equation in the diffusion approximation can be written
as follows~\cite{BFKL}
\beq
j=2-\Delta -D\,\nu ^2\,,
\eeq
where $\nu$ is related to the anomalous dimension $\gamma$ of
the twist-2 operators~\cite{FL}
\beq
\gamma =1+\frac{j-2}{2}+i\nu \,.
\eeq
The parameters $\Delta$ and $D$ are functions of the coupling constant
$\hat{a}$ and are known up to two loops. Higher order perturbative
corrections can be obtained with the use of the effective
action~\cite{eff, last}.
For large coupling constants one can expect, that the leading
Pomeron singularity in $N=4$ SUSY is moved to the point $j=2$ and
asymptotically the Pomeron coincides with the graviton Regge pole.
This assumption
is related to the AdS/CFT correspondence, formulated in the framework of
the Maldacena hypothesis claiming, that $N=4$ SUSY is equivalent 
to the superstring
model living on the 10-dimensional
anti-de-Sitter space~\cite{Malda, GKP, W}. Therefore it is
natural to impose on the BFKL equation in the diffusion approximation
the physical condition, that for the conserved energy-momentum tensor
$\theta _{\mu \nu}(x)$ having $j=2$ the anomalous dimension $\gamma$ is
zero. As a result, we obtain, that
the parameters $\Delta$ and $D$ coincide~\cite{KLOV}. In this case one
can solve
the above BFKL equation for $\gamma$
\beq
\gamma
=(j-2)\left(\frac{1}{2}-\frac{1/\Delta}{1+\sqrt{1+(j-2)/\Delta}}\right)\,.
\eeq
Using the dictionary developed in the framework of the AdS/CFT
correspondence~\cite{GKP}, one can rewrite the eigenvalue relation
for the BFKL kernel in the form
of the graviton Regge trajectory~\cite{KLOV}
\beq
j=2+\frac{\alpha '}{2}\,t\,,\,\,t=E^2/R^2\,,\,\,\alpha
'=\frac{R^2}{2}\,\Delta \,.
\eeq
On the other hand, Gubser, Klebanov and Polyakov predicted the following
asymptotics of the anomalous dimension at large $\hat{a}$ and
$j$~\cite{GKP2}
\beq
\gamma _{|\hat{a},j \rightarrow \infty} = - \sqrt{j-2}\,\Delta
^{-1/2}_{|j \rightarrow  \infty }= \sqrt{2\pi j}\, \hat{a}^{1/4}\,.
\eeq
As a result, one can obtain the explicit expression for the Pomeron intercept
at large coupling constants~\cite{KLOV, Polch}
\beq
j=2-\Delta \,,\,\,\Delta =\frac{1}{2\pi} \,\hat{a}^{-1/2}\,.
\eeq

Note, that in Ref.~\cite{L4} it was argued, that for $N=4$
SUSY the evolution equations for anomalous dimensions of
quasi-partonic operators are integrable in LLA. Later such
integrability was generalized to other operators~\cite{MZ}
and to higher loops~\cite{BS}. Using additionally the maximal
transcendentality hypothesis the integral equation
for the so-called casp anomalous dimension was constructed
in all orders of perturbation theory~\cite{ES,BES}. Further,
the anomalous dimension of twist-2 operators in four loops 
was calculated~\cite{KLRSV}, but due to the absence of so-called
wrapping contributions in the asymptotic Bethe anzatz the obtained 
results do not agree with the BFKL predictions~\cite{trajN4, KL}.

\section{Bern-Dixon-Snirnov scattering amplitudes in $N=4$ SUSY}
To calculate higher order corrections to the BFKL equation 
in QCD and supersymmetric models one should
know production amplitudes in higher orders of perturbation theory. 
Several years ago Bern, Dixon and Smirnov suggested a simple anzatz for 
the multi-gluon scattering amplitude with the maximal helicity violation
in the planar limit 
$\alpha N_c \sim 1$ for the $N=4$ super-symmetric gauge theory~\cite{BDS}.
It turns out, that this amplitude is proportional to its Born expression.
The proportionality coefficient $M_n$ for $n$ external particles 
is a function of relativistic
invariants and can be written as follows
\beq
\ln M_n= \sum _{l=1}^\infty a^l\left(f^{(l)}(\epsilon )\,
\left(\hat{I}_n^{(l)}(l\epsilon )+F_n^{(1)}(0)\right)+C^{(l)}+E_n^{(l)}(\epsilon )
\right),
\eeq
\beq
f^{(l)}(\epsilon )=f_0^{(l)}+\epsilon f_1^{(l)}+\epsilon ^2 f_2^{(l)}\,,\,\,
\gamma (a)=4\sum _{l=1}^\infty a^lf_0^{(l)}\,,\,\,\beta (a)=
\sum _{l=1}^\infty a^lf_1^{(l)}\,,\,\,\delta (a)=\sum _{l=1}^\infty 
a^lf_2^{(l)}\,, 
\eeq
where $E_n^{(1)}(\epsilon )$ can be neglected for $\epsilon \rightarrow 0$
and the constants $C^{(l)}$ and $f_r^{(l)}(\epsilon )$ are known up to rather
high order of the perturbation theory. In particular, $\gamma (a)$ is the 
so called cusp anomalous dimension which was found in all orders~\cite{ES, 
BES}
\beq
\gamma (a)=4a-4\zeta _2\,a^2+22\zeta_4\,a^3+...\,.
\eeq 
The singular function $\hat{I}_n^{(1)}(\epsilon )$ is given below
\beq
\hat{I}_n^{(1)}(\epsilon )=-\frac{1}{2\epsilon ^2}
\sum_{i=1}^n\left(\frac{\mu ^2}{-s_{i,i+1}}\right)^\epsilon
\eeq
and the finite remainders $F_n^{(1)}$ are expressed 
in terms
of logarithms and dilogarithms.
 
In Ref.~\cite{BLS} the BDS anzatz was investigated in the Regge kinematics
(see also Ref.~\cite{BNST}). 
In particular, the elastic amplitude has the Regge asymptotics
\begin{eqnarray}
M_{2\rightarrow 2}&=&\Gamma (t)\,\left(\frac{-s}{\mu ^2}\right)^{\omega (t)}\,
\Gamma (t)\, \left(1+ {\cal O}(\epsilon)\right),
\label{M2a2Regge}
\end{eqnarray}
where $\mu ^2$ is the renormalization point,
\[
\omega (t)=-\frac{\gamma (a)}{4}\,\ln \frac{-t}{\mu ^2}+\int _0^a
\frac{da'}{a'}\left(\frac{\gamma (a')}{4\epsilon}+\beta (a')\right)
\]
\begin{equation}
=\left(-\ln \frac{-t}{\mu ^2}+\frac{1}{\epsilon}\right)a+
\left[\zeta _2\left(\ln \frac{-t}{\mu ^2}-\frac{1}{2\epsilon}\right)-
\frac{\zeta _3}{2}\right]a^2 +...
\label{gluontrajectory}
\end{equation}
is the all-order gluon Regge trajectory obtained from the BDS
formula~\cite{BLS} and
\begin{eqnarray}
\ln \Gamma (t)&=&\ln  \frac{-t}{\mu ^2}\,\int _0^a
\frac{da'}{a'}\left(\frac{\gamma (a')}{8\epsilon}+\frac{\beta (a')}{2}\right)
+\frac{C(a)}{2}+\frac{\gamma (a)}{2}\,\zeta _2 \nonumber\\
&-& \int _0^a \frac{da'}{a'}
\ln \frac{a}{a'}\,\left(\frac{\gamma (a')}{4\epsilon ^2}+
\frac{\beta (a')}{\epsilon} +\delta (a')\right),
\label{ReggeonParticlevertex}
\end{eqnarray}  
is the vertex for the Reggeized gluon coupling to the external
particles. Note that the perturbative expansion for $\omega (t)$ is 
in an agreement with its direct calculations done initially in the
$\widetilde{MS}$-scheme~\cite{trajN4}.

One can 
verify that in all physical regions the BDS amplitude for one gluon
production
in the multi-Regge kinematics can be obtained with the use of
an analytic continuation from the expression~\cite{BLS}
\begin{eqnarray}
\frac{M_{2\rightarrow 3}}{\Gamma (t_1)\Gamma (t_2)} ~=~
\left(\frac{-s_1}{\mu^2}\right)^{\omega (t_1)-\omega (t_2)}
\left(\frac{-s\kappa }{\mu^4}\right)^{\omega (t_2)}c_1+
\left(\frac{-s_2}{\mu^2}\right)^{\omega (t_2)-\omega (t_1)}
\left(\frac{-s\kappa }{\mu^4}\right)^{\omega (t_1)}c_2\,,
\label{Mdosatres}
\end{eqnarray}
where  the coefficients $c_i$ are real 
\begin{eqnarray}
c_1(\kappa )&=&|\Gamma (t_2,t_1, \ln {-\kappa} )|\,
\frac{\sin \pi (\omega (t_1)-\phi _\Gamma )}{\sin \pi (\omega (t_1)-\omega (t_2))}\,,
\label{ces1}\\
c_2(\kappa )&=&|\Gamma (t_2,t_1, \ln {-\kappa} )|\,
\frac{\sin \pi (\omega (t_2)-\phi _\Gamma )}{\sin \pi (\omega (t_2)-\omega (t_1))}\,.
\label{ces2}
\end{eqnarray}
Here $\phi _\Gamma$ is the phase of the Reggeon-Reggeon-gluon vertex
$\Gamma$, {\it i.e.}
\begin{equation}
\Gamma (t_2,t_1, \ln \kappa -i\pi )=|\Gamma (t_2,t_1, \ln -\kappa )|\,
e^{i\pi \phi _\Gamma}\,,
\label{phigamma}
\end{equation}
defined by the expression
\begin{eqnarray}
\ln \Gamma (t_2,t_1, \ln -\kappa )&=&-\frac{\gamma (a)}{16}\,
\ln ^2\frac{-\kappa }{\mu ^2}-\frac{1}{2}\int _0^a \frac{da'}{a'}
\ln \frac{a}{a'}\,\left(\frac{\gamma (a')}{4\epsilon ^2}+\frac{\beta (a')}{\epsilon}
+\delta (a')\right)\nonumber\\
&&\hspace{-4cm} -\frac{\gamma (a)}{16}\ln ^2\frac{-t_1}{-t_2}-
\frac{\gamma (a)}{16}\zeta _2-\frac{1}{2}
\left(\omega (t_1)+\omega (t_2)-
\int _0^a\frac{da'}{a'}\,\left(\frac{\gamma (a')}{4\epsilon}+\beta(a')\right)\right)
\ln \frac{-\kappa }{\mu ^2}
\,.
\label{Gammavertex}
\end{eqnarray}

In the above dispersion-type representation  for the production amplitude we can use
the reality condition for the produced gluon
\begin{equation}
\kappa \rightarrow \frac{s_1s_2}{s}= \vec{k}^2_\perp \,,
\end{equation}
where $\vec{k}_\perp$ is the transverse component of its momentum
($k_\perp p_A =k_\perp p_B=0$).

In a similar way two gluon production amplitude in the multi-Regge kinematics
almost in all physical regions can be obtained by an analytic continuation
from the following dispersion-like representation for the BDS expression
\begin{eqnarray}
\frac{M_{2\rightarrow 4}}{\Gamma (t_1)\Gamma (t_3)} &=&
\left(\frac{-s_1}{\mu^2}\right)^{\omega (t_1)-\omega (t_2)}   
\left(\frac{-s_{012}\kappa _{12}}{\mu^4}\right)^{\omega (t_2)-\omega (t_3)}
\left(\frac{-s\kappa_{12} \kappa_{23}}{\mu^6}\right)^{\omega (t_3)}\,   
d_1 \nonumber\\
&+&\left(\frac{-s_3}{\mu^2}\right)^{\omega (t_3)-\omega (t_2)}
\left(\frac{-s_{123}\kappa _{23}}{\mu^4}\right)^{\omega (t_2)-\omega (t_1)}
\left(\frac{-s\kappa_{12} \kappa_{23}}{\mu^6}\right)^{\omega (t_1)}\,d_2
\nonumber\\
&+&\left(\frac{-s_2}{\mu^2}\right)^{\omega (t_2)-\omega (t_1)}
\left(\frac{-s_{012}\kappa _{12}}{\mu^4}\right)^{\omega (t_1)-\omega (t_3)}
\left(\frac{-s\kappa_{12} \kappa_{23}}{\mu^6}\right)^{\omega (t_3)}\,
d_3  \nonumber\\
&+&\left(\frac{-s_2}{\mu^2}\right)^{\omega (t_2)-\omega (t_3)}
\left(\frac{-s_{123}\kappa _{23}}{\mu^4}\right)^{\omega (t_3)-\omega (t_1)}
\left(\frac{-s\kappa_{12} \kappa_{23}}{\mu^6}\right)^{\omega (t_1)}\,d_4
\nonumber\\
&+&\left(\frac{-s_3}{\mu^2}\right)^{\omega (t_3)-\omega (t_2)}
\left(\frac{-s_1}{\mu^2}\right)^{\omega (t_1)-\omega (t_2)}
\left(\frac{-s\kappa_{12} \kappa_{23}}{\mu^6}\right)^{\omega (t_2)}\,d_5
\label{Mdosacuatro}
\end{eqnarray}
with the real coefficients $d_{i=1,2,3,4,5}$ satisfying the
relations
\begin{eqnarray}
d_1 &=& c_1(t_2,t_1,\kappa _{12})\,c_1((t_3,t_2,\kappa _{23})\,, 
\nonumber\\   
d_2 &=& c_2((t_2,t_1,\kappa _{12})\,c_2((t_3,t_2,\kappa _{23})\,, 
\nonumber\\
d_3 + d_4 &=& c_2 ((t_2,t_1,\kappa _{12})\,c_1((t_3,t_2,\kappa _{23})
\,, \nonumber\\
d_5 &=& c_1 ((t_2,t_1,\kappa _{12})\,c_2((t_3,t_2,\kappa _{23})\,,
\end{eqnarray}
where 
\[
\kappa _{12}=(\vec{q}_1-\vec{q}_2)_\perp ^2\,,\,\,
\kappa _{23}=(\vec{q}_2-\vec{q}_3)_\perp ^2\,.
\]

However, in the physical kinematical region, where $s,s_2>0$ but
$s_1,s_3<0$ the Regge factorization for the BDS amplitude is broken 
\begin{eqnarray}
&&\hspace{-1cm}
\frac{M_{2\rightarrow 4}}{\Gamma (t_1)\Gamma (t_3)} ~=~ \nonumber\\
&&\hspace{-0.8cm}C \left(\frac{-s_1}{\mu ^2}\right)^{\omega (t_1)}
\Gamma (t_2,t_1,\ln \kappa _{12}-i\pi )
\left(\frac{-s_2}{\mu ^2}\right)^{\omega (t_2)}
\Gamma (t_3,t_2, \ln \kappa _{23}-i\pi )
\left(\frac{-s_3}{\mu^2}\right)^{\omega (t_3)},
\label{24mixedregion}
\end{eqnarray}
where the coefficient $C$ is given below
\begin{equation}
C=\exp \left[\frac{\gamma _K(a)}{4} \,i\pi \,\left(
\ln \frac{\vec{q}_1^2\vec{q}_3^2}{(\vec{k}_1+\vec{k}_2)^2\mu ^2}   
-\frac{1}{\epsilon}\right)\right].
\label{coeffC}
\end{equation}
Similarly for the BDS amplitude describing the transition
$3\rightarrow 3$ in the physical region, where $s,s_2=t'_2>0$ but
$s_1,s_3<0$ we obtain the result
\begin{eqnarray}
&&\hspace{-1cm}\frac{M_{3\rightarrow 3}}{\Gamma (t_1)\Gamma (t_3)} ~=~
\nonumber\\   
&&\hspace{-0.8cm}
C' \, \left(\frac{-s_1}{\mu ^2}\right)^{\omega (t_1)}
\Gamma (t_2,t_1,\ln \kappa _{12}+i \pi )
\left(\frac{-s_2}{\mu ^2}\right)^{\omega (t_2)}
\Gamma (t_2,t_1, \ln \kappa _{23}+i\pi )
\left(\frac{-s_3}{\mu^2}\right)^{\omega (t_3)},
\end{eqnarray}
where the phase factor $C'$ is
\begin{equation}
C'=\exp \left[\frac{\gamma _K(a)}{4} \,(-i\pi ) \,
\ln \frac{(\vec{q}_1-\vec{q}_2)^2\,
(\vec{q}_2-\vec{q}_3)^2}{(\vec{q}_1+\vec{q}_3-\vec{q}_2)^2\,\vec{q}_2^2}
\right],
\label{coeffC'}
\end{equation}
which also contradicts the Regge factorization. The reason for these drawbacks
is that just in these kinematical regions the amplitudes $A_{2\rightarrow 4}$ and
$A_{3\rightarrow 3}$ should contain the Mandelstam cuts in the $j$-pane of the 
$t_2$-chanel~\cite{BLS}. 
Therefore the BDS amplitudes for these processes are not correct beyond 1 loop.
  
\section{Mandelstam cuts in the adjoint representation at LLA}

The Mandelstam cuts in the elastic amplitude appear only in the non-planar 
diagrams because the integrals for the Sudakov variables $\alpha =2kP_A/s$ 
and $\beta =2kp_B$ of the reggeon momenta $k$ and $q-k$ should have the
singularities above and below the corresponding integration contours. 
For the case of the planar diagrams this Mandelstam condition is fulfilled
only for inelastic amplitudes starting from six external particles in
the kinematical region where $s,s_2>0$ and $s_1,s_3<0$. Two reggeons in the
$t_2$-channel with an adjoint representation of the gauge group $SU(N_c)$ 
can also scatter each from 
another. The corresponding contribution to the  imaginary part in the 
$s_2$-channel for the amplitude $A_{2\rightarrow 4}$ can be written 
as follows~\cite{BLS}
\beq
\frac{1}{\pi}\Im _{s_2} M_{2\rightarrow 4}=s_2^{\omega (t_2)}\,
\int _{\sigma -i\infty}^{\sigma +i\infty}
\frac{d\omega}{2\pi i}
\left(\frac{s_2}{\mu ^2}\right)^\omega \,
\widetilde{f}_2(\omega )
\eeq
where the reduced partial wave $\widetilde{f}_2(\omega)$ is given by
\beq
\label{f-reduced}
\widetilde{f}_2(\omega )=\hat{\alpha}_\epsilon \,
|q_2|^2\int
d^{2-2\epsilon}k\,d^{2-2\epsilon}k'\,
\Phi _1 (k,q_2,q_1)\,
G_{\omega}(k,k',q_2)\,
\Phi_3 (k',q_2,q_3)\,.
\eeq
Here $\Phi _{1,2}$ are impact factors
\beq
\label{phi3}
\Phi_1(k,q_2,q_1)= \frac{k_1^* 
(q_2-k)^*}{q_2^*(k+k_1)^*}\,,\,\,
\Phi_3(k',q_2,q_3)=\frac{k_2(k'-q_2)}{q_2(k'-k_2)}\,.
\eeq

The Green's function $G_{\omega}(k,k',q_2)$
satisfies the BFKL-type equation
\begin{eqnarray}  
\omega G_{\omega}^{(8_A)}(k,k',q_2) &=& \frac{(2\pi)^3
\delta^{(2)}(k-k')}{k^2 (k+q_2)^2}
+ \frac{1}{k^2 (k+q_2)^2} \left(K^{(8_A)} \otimes G_{\omega}^{(8_A)} \right)
(k,k',q_2)\,,
\label{BFKLoctet}
\end{eqnarray}
where
\[
K^{(8_A)}(k,k';q_2)
\]
\beq
= \delta^{(2)}(k-k') \left( \omega(-|k|^2) +
\omega(-|q_2-k|^2)-2\omega (-|q|^2)
\right) +  \frac{a}{2} \frac{k^* (q_2-k) k'(q_2-k')^* + c.c.}{|k-k|^2}\,.
\eeq

The infrared divergencies are extracted in the form of the Regge factor 
$s_2^{\omega (t_2)}$ and coincide with those of the BDS amplitude, as it
should be. The partial wave $\widetilde{f}_2(\omega)$ contains the 
infrared divergency only in
the one loop
\beqn
\label{C-phase}
\hat{\alpha}_\epsilon \,
|q_2|^2\int
d^{2-2\epsilon}k\,
\frac{k^*q_1^*}{q_2^* (k+k_1)^*}\,\,\,
\frac{1}{|k|^2 |q_2 -k|^2}\,\,\,
\frac{k q_3}{q_2 (k-k_2)} =
\frac{a}{2} \left( \ln \frac{|q_1|^2 |q_3|^2}{|k_1+k_2|^2 \mu^2}
- \frac{1}{\epsilon}\right)\,,
\eeqn
which is also compatible with the BDS result. But in the upper loops
the iteration of the above equation leads to terms which are
absent in the BDS amplitude. For example, in two loops we
obtain for the imaginary part of $A_{2\rightarrow 4}$ in the 
$s_2$-channel the following expression 
\beq
A_{s_2}=\frac{a^2}{2}\,\ln s_2 \,\ln \frac{|q_1-q_3|^2|q_2|^2}{|q_1|^2|k_2|^2}\,
\ln \frac{|q_1-q_3|^2|q_2|^2}{|q_3|^2|k_1|^2}\,.
\eeq
It is symmetric with respect to the simultaneous transmutation
\beq
k_1 \leftrightarrow k_2\,,\,\,
q_1 \leftrightarrow -q_3\,.
\eeq
The same expression is valid for the imaginary part in the $s$-channel.

In a similar way we can calculate the $s$-channel imaginary part of
the amplitude
for the transition $3\rightarrow 3$
\beq
A_{s}^{3\rightarrow 3}=\frac{a^2}{2}\,\ln t'_2\,\ln \frac{|q_2-q_1-q_3|^2|q_2|^2}{|k_1|^2|k_2|^2}\,
\ln \frac{|q_2-q_1-q_3|^2|q_2|^2}{|q_3|^2|q_1|^2}\,.
\eeq
Moreover, the BFKL equation for the state with adjoint quantum numbers can be
solved explicitely and we obtain for the imarginary part in $s_2$-channel~\cite{BLS2}
\beq
\Im M_{2\rightarrow 4}\sim \sum _{n=-\infty}^\infty
\int _{-\infty}^\infty \frac{d\nu }{\nu ^2+\frac{n^2}{4}}\,  
\left(\frac{q_3^*k^*_1}{k^*_2q_1^*}\right)^{i\nu -\frac{n}{2}}\,
\left(\frac{q_3k_1}{k_2q_1}\right)^{i\nu +\frac{n}{2}}\,
\exp \left(\omega (\nu , n)\,\ln s_2\right)\,,
\eeq
where the eigenvalue of the reduced BFKL kernel for the adjoint
representation is
\beq
\omega (\nu , n) =-a\left( \psi (i\nu +\frac{|n|}{2})+
\psi (-i\nu +\frac{|n|}{2})-2\psi (1)\right)\,.
\eeq
It turns out, that the leading singularity of the $t_2$-partial wave
corresponds to $n=1$ and is situated at 
\[
j-1=\omega (t)+a(4\ln 2-2).
\]

\section{Multi-reggeon Mandelstam cuts}

Let us consider now the Mandelstam cuts constructed from
several reggeons~\cite{intBDS}. The non-vanishing contribution from
the exchange of $r+1$ reggeons appears in the planar diagrams only if the
number of the external lines is $n\ge 2r+4$. For the inelastic
transition $2\rightarrow 2+2r$ with the initial and final momenta
$p_A,p_B$ and $p_{A'},k_1,k_2,...,k_{2r},p_{B'}$, respectively, 
(see Fig 2) the cut
exists in the crossing channel with the momentum
\beq
q=p_A-p_{A'}-\sum _{l=1}^rk_l=p_{B'}-p_{B}+\sum _{l=r+1}^{2r}k_l=
\sum _{l=1}^{r+1}q'_l \,,
\eeq
where $q'_l$ are momenta of $r+1$-reggeons.
The corresponding amplitude has the form
\beq
A_{2\rightarrow 2+2r}\sim
\int \frac{d^2q'_1d^2q'_2...d^2q'_{r}}{(2\pi )^{r}\,s^r}
\,\prod_{l=1}^{r+1}\frac{(-s)^{j(-\vec{q'}_l^2)}}{|q'_l|^2}\Phi _1
(\vec{q}'_1,...,\vec{q}'_{r+1})\Phi _2 (\vec{q}'_1,...,\vec{q}'_{r+1})\,.
\eeq
The impact factors $\Phi _{1,2}$ are given in terms of the integrals over
the Sudakov parameters $\alpha '_l=2q'_lp_A/s,\,\beta' _l=2q'_lp_B/s$ from
the reggeon-particle scattering amplitudes $f_{1,2}$
\beq
\Phi _1=\prod _{l=1}^{r-1}\int _L\frac{s\,d\alpha '_l}{2\pi i}\,f_1\,,\,\,
\Phi _2=\prod _{l=1}^{r-1}\int _L\frac{s\,d\beta '_l}{2\pi i}\,f_2\,.
\eeq

The tree expressions for the amplitudes $f_{1,2}$ appearing
in the planar diagrams in QCD are given below
\[
\frac{f_1}{I_1}=\frac{1}{(p_A-q'_1)^2}\frac{1}{(p_A-k_0-q'_1)^2}...
\frac{1}{(p_A-\sum _{l=1}^{r}q'_l-\sum _{l=0}^{r-2}k_l)^2}
\frac{1}{(p_A-\sum _{l=1}^{r}q'_l-\sum _{l=0}^{r-1}k_l)^2}\,,
\]
\[
\frac{f_2}{I_2}=\frac{1}{(p_B+q'_1)^2}\frac{1}{(p_B-k_{2r+1}+q'_1)^2}...
\frac{1}{(p_B+\sum _{l=1}^{r}q'_l-\sum _{l=r+3}^{2r+1}k_l)^2}
\frac{1}{(p_B+\sum _{l=1}^{r}q'_l-\sum _{l=r+2}^{2r+1}k_l)^2}\,,
\]
where $k_0=p_{A'},\,k_{2r+1}=p_{B'}$. The additional factors $I_{1,2}$
contain effective reggeon vertices for the production and scattering of the
gluons with the same helicity. They can be written
in the multi-Regge kinematics as follows
\[
I_1=\prod _{l=1}^r
\frac{q^{\prime *}_{l+1}(Q-\sum _{t=1}^lq^{\prime }_t
-\sum _{t=1}^{l-1}k_t)}{(Q^*-\sum _{t=1}^{l+1}q^{\prime *}_t
-\sum _{t=1}^{l-1}k_t^*)}\,\prod _{l=1}^r\beta _r\,,
\]
\[
I_2=\prod _{l=1}^r
\frac{q^{\prime }_{l+1}(\widetilde{Q}^*+\sum _{t=1}^lq^{\prime *}_t
-\sum _{t=1}^{l-1}k^*_{2r-t+1})}{(\widetilde{Q}+\sum _{t=1}^{l+1}q^{\prime }_t
-\sum _{t=1}^{l-1}k_{2r-t+1})}\,\prod _{l=1}^r\alpha _r\,,
\]
where $Q=p_A-p_{A'},\,\widetilde{Q}=p_B-p_{B'}$ and the Sudakov variables
of the produced particles
$\alpha _l=2k_lp_A/s,\,\beta _l=2k_lp_B/s$
are strongly ordered
\[
1\gg |\beta _1| \gg |\beta _2|...\gg |\beta _{2k}|\,,\,\,|\alpha _1|\ll
|\alpha _2|\ll ...
|\alpha _{2k}|\ll 1\,.
\]
In the physical
region, where the signs of the Sudakov parameters of momenta $k_l$ alternate with
the index $l$
\[
\beta _1,\,\alpha _{2r}<0\,;\,\,\beta _2,\,\alpha _{2r-1}>0
\,;\,\,\beta _3,\,\alpha _{2r-2}<0\,;...\,,
\]
which is equivalent to the following constraints on the invariants
\beq
s_1<0,s_2<0,...,s_r<0,s_{r+1}>0,s_{r+2}<0,s_{r+3}<0,...,s_{2r+1}<0,s>0\,,
\eeq
the integrands in expressions for $\Phi _{1,2}$ contain poles
over the variables $\alpha' _l,\,\beta '_l$ above and below the
integration contours $L$ over. Therefore $\Phi _{1,2}$
are non-zero and
can be calculated by taking residues from
the poles 
\beq
\Phi _1(\vec{q}'_1,...,\vec{q}'_{r+1})=\prod _{l=1}^r
\frac{q^{\prime *}_{l+1}}{(Q^*-\sum _{s=1}^lq^{\prime *}_s-\sum _{s=1}^{l-1}k^*_s)
\,(Q^*-\sum _{t=1}^{l+1}q^{\prime *}_t
-\sum _{t=1}^{l-1}k_t^*)}\,,
\eeq
\beq
\Phi _2(\vec{q}'_1,...,\vec{q}'_{r+1})=  
\prod _{l=1}^r\frac{q'_{l+1}}{(\widetilde{Q}+\sum _{s=1}^lq'_s-
\sum _{s=1}^{l-1}k_{2r -s+1})\,(\widetilde{Q}+\sum _{t=1}^{l+1}q^{\prime }_t
-\sum _{t=1}^{l-1}k_{2r-t+1})}\,.
\eeq

In the case of production of $2r$ gluons with the same helicity
the amplitude in $N=4$ SUSY is proportional to the Born expression.
In the leading logarithmic approximation 
for the $r+1$-reggeon contribution to the $s_{r+1}$-channel the 
proportionality factor has the form
\beq
f_{LLA}^{2\rightarrow 2+2r}=\left(i\,\frac{g^2\,N_c}{4\pi}\right)^{r}\,  
Q^*\widetilde{Q}\int \prod _{l=1}^r\frac{\mu ^{2\epsilon}d^{2-2\epsilon }
p_l}{(2\pi)^{1-2\epsilon}}\,\frac{\mu ^{2\epsilon}d^{2-2\epsilon }
p'_l}{(2\pi)^{1-2\epsilon}}\,\prod _{l=1}^{r}
\frac{k^*_lk_{2r-l}}{|p_l|^2}\,\frac{G(p,p';s_{r+1})}{|p_{r+1}|^2}\Phi _1\,
\Phi_2\,,
\eeq
where we introduce the new notation $p_l$ for the reggeon momenta $q'_l$. The
multi-reggeon Green function satisfies the equation~\cite{intBDS}
\beq
\frac{\partial}{\partial \ln s_{r+1}}\,
G(\vec{p},\vec{p}';s_{r+1})=
K\,G(\vec{p},\vec{p}';s_{r+1})\,,\,\,
G(\vec{p},\vec{p}';0)=
\prod _{l=1}^r\frac{(2\pi )^{1-2\epsilon}}{\mu ^{2\epsilon}}\,
\delta ^{2-2\epsilon}(p_{l}-p'_{l})\,.
\eeq
Here the kernel $K$ in LLA can be expressed in terms of
the infraredly stable Hamiltonian $H$ 
\beq
K=\omega (t) -\frac{g^2N_c}{16 \pi ^2}\,H\,,\,\,
\omega (t)=a\left(\frac{1}{\epsilon}-\ln \frac{-t}{\mu _2}\right)\,,
\,\,t=-|q|^2\,,
\eeq
\beq
H=\ln \frac{|p_1|^2|p_{r+1}|^2}{|q|^4}+\sum _{l=1}^{r}H_{l,l+1}\,,
\eeq
where the pair Hamiltonian is
\beq
H_{l,l+1}=\ln |p_l|^2+\ln |p_{l+1}|^2+
p_l\,p_{l+1}^*\,\ln |\rho _{l,l+1}|^2\,\frac{1}{p_l\,p_{l+1}^*}+
p_l^*\,p_{l+1}\,\ln |\rho _{l,l+1}|^2\,\frac{1}{p_l^*\,p_{l+1}}\,.
\eeq

\section{Integrable open Heisenberg spin chain}
The Hamiltonian for the gluon composite state 
in the ajoint representation has the property of the holomorphic
separability~\cite{intBDS}
\beq
H=h+h^*\,,\,\,h=\ln \frac{p_1\,p_{r+1}}{q^2}+
\sum _{l=1}^{r}h_{l,l+1}\,,
\eeq
where
\beq
h_{l,l+1}=\ln p_l+\ln p_{l+1}+
p_l\,\ln \rho _{l,l+1}\,\frac{1}{p_l}+
p_{l+1}\,\ln \rho _{l,l+1}\,\frac{1}{p_{l+1}}\,.
\eeq

Using the duality transformations (cf.~\cite{dual})
\begin{equation}
p_1=z_{0,1}\,,\,\,p_r=z_{r-1,r}\,,\,\,q=z_{0,n}\,,\,\,\rho _{r,r+1}=
i\frac{\partial}{\partial z_r}=i\partial _r\,,
\end{equation}
the holomorphic hamiltonian can be rewritten as follows
\begin{equation}
h=\ln \frac{z_{0,1}\,z_{n-1,n}}{z_{0,n}^{2}}
+\sum_{r=1}^{n-1}h_{r,r+1}\,,
\label{hol1}
\end{equation}
where
\begin{equation}
h_{r,r+1}=2\ln (\partial _r)
+\frac{1}{\partial _r}\,\frac{1}{z_{r-1,r}}+
\frac{1}{\partial _r}\,\frac{1}{z_{r+1,r}}++2\gamma \,.
\end{equation}
Here and later we neglect the pure imaginary contribution $2\ln (i)$
because it is cancelled in the total hamiltonian $H$. 

One can verify, that in the new variables  $h$ is invariant 
under the M\"{o}bius
transformations
\begin{equation}
z_k\rightarrow \frac{az_k+b}{cz_k+d}\,.
\end{equation}
Therefore we can put
\begin{equation}
z_0=0\,,\,\,z_n=\infty \,,
\end{equation}
Further, by regrouping the terms one can write the holomorphic
hamiltonian for $n$-reggeon interactions in the adjoint
representation in other form~\cite{intBDS}
\begin{equation}
h=-2\ln z_{0,n}
+\ln (z_{0,1}^2\partial _1)+
\ln (z_{n-1,n}^2\partial _{n-1})+2\gamma +
\sum_{r=1}^{n-2}\,h'_{r,r+1}\,,
\end{equation}
where
\[
h'_{r,r+1}=\ln (z_{r,r+1}^2\partial _r)+
\ln (z_{r,r+1}^2\partial _{r+1})-2\ln z_{r,r+1}+
2\gamma
\]
\begin{equation}
=\ln (\partial _r)+\ln (\partial _{r+1})
+\frac{1}{\partial _r}\,\ln z_{r,r+1}\,\partial _r+
\frac{1}{\partial _{r+1}}\,\ln z_{r,r+1}\,\partial _{r+1}
+2\gamma \,.
\end{equation}
The pair hamiltonian $h'_{r,r+1}$ coincides in fact with
the expression (\ref{pairh}) in the coordinate representation
acting on the wave function with 
non-amputated propagators.

The remarkable property of $h$ is 
its commutation with the matrix element $D(u)$ of the monodromy matrix 
(\ref{ABCD}) introduced above for the description of the integrability 
of the BKP equations in the multi-color QCD~\cite{intBDS}
\beq
[D(u),h]=0\,.
\eeq
Therefore if we write $D(u)$ as a polynomial in $u$
\begin{equation}
D(u)=\sum _{k=0}^{n-1}u^{n-1-k}\,q'_k\,,
\end{equation}
then the differential operators
\begin{equation}
q'_0=1\,,\,\,q'_{1}=-i\sum _{r=1}^{n-1}z_r\,
\partial _r \,,
\end{equation}   
\begin{equation}
q'_k=-\sum _{0<r_1<r_2<...<r_{k}<n}z_{r_1}\,
\prod _{s=1}^{k-1}z _{r_s,r_{s+1}}\,
\prod _{t=1}^k i\partial _{r_t}
\end{equation}
are independent integrals of motion with the properties
\beq
[q'_k,h]=[q'_k,q'_t]=0\,.
\eeq 
It turns out, that $h$
coincides with the local hamiltonian
of the open integrable Heisenberg model in which spins
are generators of the M\"{o}bius group. 

To solve this model one can use the algebraic Bethe anzatz.
In this case it is convenient to go to the transposed space,
where there exists the pseudo-vacuum state $\Psi _0$
\beq
\Psi _0=\prod_{r=1}^{n-1}z_r^{-2}\,,
\eeq
satisfying the equation
\beq
C^t(u)\Psi _0=0\,.
\eeq
Here $C^t(u)$ is the transposed matrix element $C(t)$ of the
monodromy matrix (\ref{ABCD}). The eigenvalues of the hamiltonian and 
the integral of motion $D(u)$ are constructed by applying
the product of its matrix elements $B^t(u)$ to the pseudovacuum
state
\beq
\Psi _k=\prod _{r=1}^k B^t(u_r)\,\Psi _0\,.
\eeq
For such eigenfunctions the spectral parameters $u_r$ should obey 
so-called Bethe equations. Instead one can introduce the Baxter function
which is the generating function of the Bethe roots
\beq
Q(u)=\prod _{k=1}^\infty (u-u_k)\,.
\eeq 
Generally the number of the roots $u_k$ is infinite. The Baxter
function satisfies the Baxter equation which is reduced to the simple
recurrent relation for our open spin chain
\begin{equation}
\Lambda (u)\,Q(u)=(u+i)^{n-1}\,Q(u+i)\,,
\end{equation}
where $\Lambda (u)$ is an eigenvalue of the integral of motion $D(u)$
and can be written in terms of its roots
\begin{equation}
D(u)\,\Psi _{a_1,a_2,...,a_{n-1}}=\Lambda (u)\,\Psi _{a_1,a_2,...,a_{n-1}}
\,,\,\, \Lambda (u)=\prod _{r=1}^{n-1}(u-ia_r)\,.
\end{equation}
As a result, the solution of the Baxter equation can be found
in the form~\cite{intBDS}
\beq
Q(u)=\prod _{r=1}^{n-1}\frac{\Gamma (-iu-a_r)}{\Gamma (-iu+1)}
\eeq
up to a possible factor being a periodic function of $-iu$.

The Regge trajectory of the composite state  of $n-1$ gluons has the 
additivity property
\begin{equation}
\omega _n(t)=\omega (t)-\frac{a}{2}\,E\,,\,\,
E=\epsilon +\widetilde{\epsilon }\,,
\end{equation}
\begin{equation}
\epsilon =
\sum _{r=1}^{n-1}\epsilon (a_r)\,,\,\,\widetilde{\epsilon}=
\sum _{r=1}^{n-1}\epsilon (\widetilde{a}_r)\,,
\end{equation}
where 
\beq
\epsilon (\widetilde{a})=\psi (a)+\psi (1-a)-2\psi (1)\,,\,\,
a_r=i\nu _r+\frac{n_r}{2}\,.
\eeq

\section{Three gluon composite state}

The wave funcion of the 
three gluon composite state in the adjoint representation can be
constructed as a bilinear combination of eigenfunctions of
the integrals of motion $D(u)$ and $D^*(u)$ having the property
of single-valuedness in the coordinate space~\cite{intBDS}   
\begin{equation}
\Psi \sim z^{a_1+a_2}_2\,(z^*_2)^{\widetilde{a_1}+\widetilde{a_2}}\,
\int \frac{d^2y}{|y|^2}\,
y^{-a_2}(y^*)^{-\widetilde{a_2}}\,\left(\frac{y-1}{y-x}\right)^{a_1}\,
\left(\frac{y^*-1}{y^*-x^*}\right)^{\widetilde{a_1}}\,,\,\,
x=\frac{z_2}{z_1}\,.
\end{equation}

One can perform its Fourie transformation to the momentum space
\begin{equation}
\Psi ^t(\vec{p}_1,\vec{p}_2 )=(p_1+p_2)^{-a_1-a_2}
(p_1^*+p_2^*)^{-\widetilde{a}_1-\widetilde{a}_2}
\,\phi (\vec{y})\,,\,\,y=\frac{p_2}{p_1}\,,
\end{equation}
where
\begin{equation}
\phi (\vec{y})=\int d^2t\,
\left(\frac{1}{t\,y}+1\right)^{a_1}\,
\left(\frac{1}{t^*\,y^*}+1\right)^{\widetilde{a}_1}\,
(1-t)^{a_2-1}\,(1-t^*)^{\widetilde{a}_2-1}\,.
\end{equation}
This function can be presented in terms of its Mellin transformation
\begin{equation}
\Psi ^t(\vec{p}_1,\vec{p}_2 )=(p_1+p_2)^{-a_1-a_2}
(p_1^*+p_2^*)^{-\widetilde{a}_1-\widetilde{a}_2}
\,\int d^2u\,
\phi (u,\widetilde{u})\,\left(\frac{p_1}{p_2}\right)^{-iu}\,
\left(\frac{p_1^*}{p_2^*}\right)^{-i\widetilde{u}}
\,,
\end{equation}
where
\begin{equation}
-iu=i\nu _u+\frac{N_u}{2}\,,
\,\,-i\widetilde{u}=i\nu _u-\frac{N_u}{2}\,\,,\,\,\,
\int d^2u\equiv \int _{-\infty}^{\infty} d\nu _u \sum
_{N_u=-\infty}^{\infty} \,.
\end{equation}
and
\begin{equation}
\phi (u,\widetilde{u})=
\frac{\pi ^2\Gamma (1+\widetilde{a}_1)
\Gamma (a_2)}{\Gamma (-a_1)\,\Gamma (1-\widetilde{a}_2)}\,
\frac{\Gamma (iu)\Gamma (1+i\widetilde{u})}{\Gamma (-iu)\,
\Gamma (1-i\widetilde{u})}\,\frac{\Gamma (-iu-a_1)  
\,\Gamma (-iu-a_2)}{
\Gamma (1+i\widetilde{u}+\widetilde{a}_1)
\Gamma (1+i\widetilde{u}+\widetilde{a}_2)}\,.
\end{equation}
Really the last form of $\Psi^t$ corresponds to the 
Baxter-Sklyanin representation~\cite{Veg},
because the function $\phi$ is a product of the pseudovacuum
state and the Baxter function~\cite{intBDS}
\begin{equation}
\phi (u,\widetilde{u})=u\,\widetilde{u}\,Q(u,\widetilde{u})\,,
\end{equation}
where
\begin{equation}
Q(u,\widetilde{u})\sim
\frac{\Gamma (iu)\Gamma (i\widetilde{u})}{\Gamma (1-iu)\,
\Gamma (1-i\widetilde{u})}\,\frac{\Gamma (-iu-a_1)
\,\Gamma (-iu-a_2)}{
\Gamma (1+i\widetilde{u}+\widetilde{a}_1)
\Gamma (1+i\widetilde{u}+\widetilde{a}_2)}\,.
\end{equation}

\section{Discussion of obtained results}

It was demonstated, that Pomeron in QCD is a composite state of reggeized
gluons.
The BFKL dynamics is integrable in LLA. In the next-to-leading
approximation in $N=4$ SUSY the equation for the Pomeron
wave function has remarkable properties including the analyticity in
the conformal spin $n$ and the maximal transcendentality. In this model the
BFKL Pomeron coincides with the reggeized graviton.
The BDS ansatz for scattering amplitudes in $N=4$ SUSY does not agree with 
the BFKL approach in the multi-Regge kinematics. The reason for this drawback 
is the absence of the Mandelstam cuts. The BFKL-like equation for the
composite state of two reggeized gluons with adjoint quantum numbers is 
explicitely solved. It is shown, that the equation for the composite
state of an arbitrary number of reggeized gluons in the adjoint representation
is equivalent to the Schr\"{o}dinger equation for an integrable open Heisenberg
spin chain. The wave function for three gluon composite state is constructed in the 
Baxter-Sklyanin  representation. 

In the conclusion I thank L.D. Faddeev, J. Bartels and A. Sabio Vera for
helpful discussions.

\end{document}